# High figure-of-merit in the heavy-fermion $UN_2$ system for radioisotope thermoelectric applications

Z. Z. Zhou[1], D. D. Fan[1], H. J. Liu[1,*], J. Liu[2,†]

[1]*Key Laboratory of Artificial Micro- and Nano-Structures of Ministry of Education and School of Physics and Technology, Wuhan University, Wuhan 430072, China*

[2]*Institute of Materials, China Academy of Engineering Physics, Mianyang 621907, China*

**Abstract**

The design of uranium-based thermoelectric materials presents a novel and intriguing strategy for directly converting nuclear heat into electrical power. Using high-level first-principles approach combined with accurate solution of Boltzmann transport equation, we demonstrate that a giant *n*-type power factor of 13.8 mWm$^{-1}$K$^{-2}$ and a peak *ZT* value of 2.2 can be realized in the heavy-fermion $UN_2$ compound at 700 K. Such promising thermoelectric performance arises from the large degeneracy ($N_v$ = 14) of heavy conduction band coupled with weak electron-phonon interactions, which is in principle governed by the strong Coulomb correlation among the partially filled U-5*f* electrons in the face-centered cubic structure. Collectively, our theoretical work suggests that the energetic $UN_2$ is an excellent alternative to efficient radioisotope power conversion, which also uncovers an underexplored area for thermoelectric research.

## 1. Introduction

Actinide series possess some unique and interesting physical properties owing to the partially filled 5*f* electrons, which have received wide attention as advanced nuclear fuels [1−3]. In general, the actinides are more or less radioactive. Fortunately, the uranium (U) element with low enough activity can be handled with minimal licensure requirements and the uranium nitrides have been considered as potential fuels for Generation IV reactors [4]. On the other hand, the U-based compounds have also attracted extensive interest as heavy-fermion systems for thermoelectric (TE) applications [5]. Such energetic materials could be utilized to generate power by harvesting the heat produced from the decay of radioactive isotopes, which is regarded as a key technology for nuclear heat recovery and space exploration [6].

---

[*]Author to whom correspondence should be addressed. Email: phlhj@whu.edu.cn
[†]Author to whom correspondence should be addressed. Email: jingliu@mail.ustc.edu.cn

Distinguished from the extensively investigated tetradymites, the TE properties of the U-based materials [7−10] are less studied to data, especially for the systems consisting of U and nitrogen (N) element [11,12]. Uranium nitrides include three stable phases with nominal formulas of UN, $U_2N_3$, and $UN_2$ [13]. Experimentally, Didchenko and Gortsema [7] measured the TE properties of UN and found a maximum figure of merit (*ZT*) of 0.6 at 700 K. Besides, the room temperature *ZT* value of 0.034 was reported by Samsel-Czekała *et al.* [12] for the UN with the measured lattice thermal conductivity ($\kappa_l$) of 13 $Wm^{-1}K^{-1}$. Recently, Hu *et al.* [14] suggested that the $U_2N_3$ may be potential TE material, which exhibits a largest power factor of 3.58 $\mu Wcm^{-1}K^{-2}$ at 383 K. In comparison, fewer first-principles investigations were performed on the uranium nitrides since the conventional density functional theory (DFT) could not capture the electronic localization effects caused by the strong Coulomb correlations [15,16]. Weck *et al.* [17] reported all-electron relativistic calculations of the band structures of the $UN_2$ crystal and found an indirect gap of 0.9 eV. Using a Hubbard parameter (*U*) of 2.0 eV, Lu *et al.* [18] revealed that the 5*f* electrons of the U element indeed play an important role in the band edges of the $UN_2$. Presently, there is no first-principles study on the electronic transport of the uranium nitrides, and neither theoretical nor experimental work has been found to investigate the TE properties of the $UN_2$ compound. Moreover, U−N systems with higher nitrogen concentration have been demonstrated to exhibit better corrosion resistance [19], indicating that the $UN_2$ crystal may be more desirable for the application in extreme environmental conditions. It is thus of vital importance to evaluate the TE potential of the energetic $UN_2$, and comprehensive first-principles calculations are needed to clarify the effect of the partially filled U-5*f* electrons on the transport properties.

In this work, we demonstrate through a high-level first-principles study that ultrahigh power factor and excellent TE performance can be realized in the heavy-fermion UN$_2$ compound. It is found that the partially filled 5*f* electrons of the U atoms can lead to a simultaneously large Seebeck coefficient and electrical conductivity, which gives a highest *ZT* value of 2.2 for the *n*-type system at 700 K.

## 2. Computational methods

Within the framework of DFT [20,21], the electronic band structure calculation is performed by employing the Perdew–Burke–Ernzerhof (PBE) functional [22], as implemented in the Vienna *ab-initio* simulation package [23]. The energy cutoff is 700 eV and a $25\times25\times25$ Monkhorst-Pack ***k***-mesh is used for the Brillouin zone integrations. To describe the effects of the strong Coulomb correlations, the DFT calculations are bridged with the relativistic Hartree-Fock approximation (DFT+*U*) [24−26] with the spin-orbital coupling explicitly considered. The quasiparticle self-energy corrections with the *GW* approximation of the many-body effects [27] are also included for obtaining accurate band structure. The electronic (Seebeck coefficient $S$, electrical conductivity $\sigma$, and electronic thermal conductivity $\kappa_e$) and phonon (lattice thermal conductivity $\kappa_l$) transport coefficients are computed by using the Boltzmann theory [28], as performed in the BoltzTraP code [29] and ShengBTE [30] program, respectively. Here we calculate the ***k***-resolved carrier relaxation time ($\tau_c$) by fully considering the electron-phonon coupling (EPC) based on the density functional perturbation theory [31] and Wannier interpolation techniques [32]. To ensure the convergence, we first obtain the EPC matrix element on coarse meshes of $10\times10\times10$ ***k***-points with $5\times5\times5$ ***q***-points, and then interpolate them to dense grids of $40\times40\times40$ ***k***-points with $40\times40\times40$ ***q***-points, as coded in the so-called electron-phonon Wannier package [33].

## 3. Results and discussion

Uranium nitrides could crystallize as UN$_2$ phase below 723 K [34]. Fig. 1(a) plots the unit cell of UN$_2$, which possesses a face-centered cubic (FCC) CaF$_2$-type lattice structure with space group of $Fm\bar{3}m$. As mentioned above, we adopt the DFT+$U$ methods to derive optimized lattice parameter ($a$) and accurate band structures for the system, where the introduced Hubbard parameter $U$ should be carefully tested. Fig. 1(b) plots the total energy of UN$_2$ as a function of the lattice parameter with the value of $U$ ranged from 1 ~ 5 eV. It is obvious that bigger Hubbard parameter corresponds to relatively larger lattice constant, and $U$ = 2 eV is finally adopted since the calculated $a$ = 5.31 Å exhibits perfect agreement with that reported in experimental work [13].

Fig. 2(a) shows the electronic band structures of UN$_2$ by using DFT+$U$ and considering the quasiparticle self-energy corrections. The band gap is calculated to be 0.91 eV, with the conduction band minimum (CBM) and valance band maximum (VBM) located at the **k**-points of (0.500, 0.500, 0.500) and (0.322, 0.322, 0.644), respectively. Noticeably, we see that both the conduction and valance band exhibit rather weak dispersions near the Fermi level, which lead to relatively large effective mass. Indeed, our additional calculations find that the density of state (DOS) effective mass ($m^*_{dos}$) at VBM and CBM are 1.11 $m_0$ and 1.19 $m_0$, respectively. Besides, we see there is a conduction band extremum (CBE) at the **X** points, which exhibits nearly identical energy and larger $m^*_{dos}$ (2.36 $m_0$) compared with those of CBM. Such large DOS effective mass in the heavy-fermion system is believed to be rooted from the strong Coulomb interactions among the partially filled 5$f$ electrons of the U atom. As confirmed in Fig. 2(b), the DOS around the VBM are contributed by the U-5$f$ and N-2$p$ electrons. In contrast, the CBM are dominated by the localized U-5$f$ states which yield rather higher DOS and in turn give fairly larger Seebeck coefficient according to the Mahan-Sofo theory [35]. On the other hand, we find that both the valance and conduction band edges exhibit quite large valley degeneracy ($N_v$) in the whole Brillouin zone of the FCC structure. Fig. 2(c) and 2(d) plot the isoenergy surfaces (0.1 eV from the band extremum) of the top valence band and bottom conduction band, respectively. We see clearly that the VBM has a much larger $N_v$ of 12 since the valance pocket locates at the off-symmetry points of the Brillouin zone, and the total DOS effective mass ($m_T^* = N^{2/3}m_{dos}^*$) of hole is calculated to be as high as 5.82 $m_0$. On the other hand, both the CBM and CBE are located at the high-symmetry points with respective $N_v$ of 8 and 6, leading to an ultrahigh valley degeneracy of 14 for the conduction band edge. An extremely large electron $m_T^*$ of 10.11 $m_0$ is therefore obtained with the averaged DOS effective mass calculated to be 1.74 $m_0$. The multiple degenerate valleys and consequently big $m_T^*$ could greatly enhance the carrier concentration and thus the electrical conductivity at the band edge, where the large Seebeck coefficient is still maintained [36] as demonstrated in the following discussion.

Fig. 3(a) plots the carrier relaxation time of UN$_2$ at 300 and 700 K derived from complete EPC calculations. We do not consider the results at higher temperature as UN$_2$ will decompose into U$_2$N$_3$ above 723 K [34]. It can be seen that the $\tau_c$ at 700 K is obviously smaller than that at 300 K since more carriers are populated at higher temperature leading to enhanced scattering rates. Note that smaller DOS effective mass near the band edge usually corresponds to relatively higher relaxation time due to the constricted carrier scattering phase space. However, here we find larger *n*-type relaxation time with bigger electron DOS effective mass compared with that of *p*-type system. Such unusual behavior should be ascribed to the weaker EPC strength for the *n*-type UN$_2$, which is characterized by the smaller absolute value of the deformation potential constant ($E_{DP}$) [37]. Indeed, the calculated $E_{DP}$ of *n*-type UN$_2$ (6.7 eV) is significantly lower than that of the *p*-type system (10.1 eV), which is responsible for the higher electron relaxation time and also suggests rather larger *n*-type electrical conductivity. It should be mentioned that previously reported $E_{DP}$ [37] usually ranges from 8 ~ 35 eV for typical semiconductors. The lower $E_{DP}$ of 6.7 eV thus implies the weak EPC in the heavy-fermion UN$_2$, which leads to relatively larger relaxation time compared with that of the traditional TE material Bi$_2$Te$_3$ [38]. Such heavy band with small $E_{DP}$ was also observed in other compounds such as ZrNiSn [37] and MgAgSb [39], which have been suggested to exhibit very promise TE performance. Fig. 3(b) shows the $E_{DP}$ of some typical TE materials [37,39−42] as a function of the DOS effective mass, where the results of the tetradymites are obtained from our additional calculations. It is interesting that compounds with bigger $m_{dos}^*$ tend to possess lower $E_{DP}$, which can achieve larger Seebeck coefficient and higher electrical conductivity simultaneously so that excellent TE performance of heavy-fermion systems is expected. On the other hand, the extremely heavy U atom means that larger strains are required to induce lattice deformation in the UN$_2$ compound. Consequently, the calculated elastic modulus (481 GPa) is remarkably larger than those of typical TE materials such as SnSe (58 GPa) [43] and Bi$_2$Te$_3$ (69 GPa) [38], which also greatly weaken the electron-phonon scattering in the UN$_2$ system.

Based on the Boltzmann theory, we can obtain the electronic and phonon transport coefficients of UN$_2$, as well as the figure of merit given by $ZT = S^2\sigma T/(\kappa_e + \kappa_l)$. Note that the UN$_2$ exhibits simultaneously higher electron DOS effective mass and relaxation time. We thus focus on the *n*-type system in the following discussions. In Fig. 4(a) ~ 4(d), the *ZT* value, the power factor ($S^2\sigma$), the absolute Seebeck coefficient ($|S|$), and the electrical conductivity ($\sigma$) are plotted as a function of electron concentration in the temperature range from 300 to 700 K. The black squares indicate the optimized carrier concentration where the *ZT* value is maximized. It is obvious that with increasing temperature, the optimized carrier concentration is obviously enhanced and the value at 700 K is as high as $1.1\times10^{21}$ cm$^{-3}$, which is caused by the large band degeneracy discussed above. Fortunately, elements in the early actinide series enable a variety of defects and impurities accommodated in the atomic structures owing to the wide range of valence states. It is thus quite favorable to obtain desirable carrier concentration in the UN$_2$ compound so that enhanced TE performance can be realized [6]. Indeed, we see from Fig. 4(a) that a maximum *ZT* of 2.2 is achieved at 700 K for the *n*-type UN$_2$, which should be traced back to the ultrahigh power factor (13.8 mWm$^{-1}$K$^{-2}$) shown in Fig. 4(b). Such a giant value is even higher than that of the state-of-art TE material SnSe [44]. As plotted in Fig. 4(c), the optimized Seebeck coefficient at 700 K can be as high as 300 μVK$^{-1}$, which is caused by the large DOS effective mass at the flat conduction band edge. On the other hand, we see a much higher electrical conductivity of $3.1\times10^5$ Sm$^{-1}$ in Fig. 4 (d) at the optimized carrier concentration, as the weak electron-phonon interactions and multiple degenerate valleys distinctly facilitate the electron transport. The record high power factor should thus be attributed to the simultaneously large Seebeck coefficient and electrical conductivity, which suggests an intriguing strategy of finding good TE materials containing unique actinide elements. Furthermore, it is noted that the UN$_2$ compound exhibits reasonably large *ZT* above 1.0 in a wide range of carrier concentration at 600 and 700 K. For instance, the *ZT* at 600 K can reach 1.1 when the carrier concentration is $1.0\times10^{19}$ cm$^{-3}$, which is within the doping range of typical thermoelectric materials and could be easily realized in the experiment.

Compared with those of good TE materials such as Bi$_2$Te$_3$ [38] and SnSe [44], the lattice thermal conductivity ($\kappa_l$) of UN$_2$ is much higher with the value of 11.2 (5.0) Wm$^{-1}$K$^{-1}$ at 300 (700) K. To further enhance the TE performance, one can reduce the $\kappa_l$ by using nanostructuring technique [45]. In principle, both the electrons and phonons with mean free path (MFP) longer than the characteristic length of a nanostructure can be efficiently scattered so that they do not contribute to the transport. Our additional calculations reveal that the largest MFP of electron is remarkably lower than that of phonon for the UN$_2$, as also observed in many other systems [46]. Fig. 5(a) shows the accumulative $\kappa_l$ of UN$_2$ with respect to the phonon MFP, where the vertical dashed line at each temperature indicates the maximum length of electron MFP. If the sample length can be reduced to the value marked by the vertical dashed line, the $\kappa_l$ of UN$_2$ will be decreased by more than 60%, while the electronic transport is almost unchanged. As a consequence, we see from Fig. 5(b) that the TE performance can be significantly enhanced, where the highest *ZT* value above 3.0 is realized for the *n*-type system at 700 K. By considering the temperature range from 300 to 700 K, an average figure of merit (*ZT$_{ave}$*) as high as 1.9 is obtained for the *n*-type system (the inset of Fig.

5(b)), which is much larger than those of typical good TE materials such as SnSe, MgAgSb, and PbTe [39,44,47]. In addition, it is interesting to find that the *ZT* of *p*-type system could also be improved to larger than 2.0 with a reasonably high $ZT_{ave}$ of 1.2, making the energetic $UN_2$ compound a quite plausible candidate for realistic TE applications.

## 4. Summary

We demonstrate via sophisticated first-principles calculations that extraordinary TE performance can be realized in the heavy-fermion $UN_2$ system ascribed to the cooperation of large DOS effective mass, high band degeneracy, and weak electron-phonon interaction. Note that the optimized carrier concentration could be achieved by 2 at% O/S or 1 at% F/Cl doping, which is experimentally possible [6−8] due to the wide range of valence states for the U atom. As the energetic nuclear material enabling stable and enduring heat release, the $UN_2$ can be utilized to generate electrical power in some special environments such as outer space. Our theoretical study thus uncovers a novel and significant opportunity of searching heavy-fermion systems containing unique actinide elements as radioisotope TE generators, which is also of crucial importance for military application and planetary exploration.

**Acknowledgments**

We are grateful for financial support from the National Natural Science Foundation of China (Grants No. 51772220 and No. 11574236), and the Fund of Science and Technology on Surface Physics and Chemistry Laboratory (Grant No. 6142A02180404). The numerical calculations in this work have been done on the platform in the Supercomputing Center of Wuhan University.

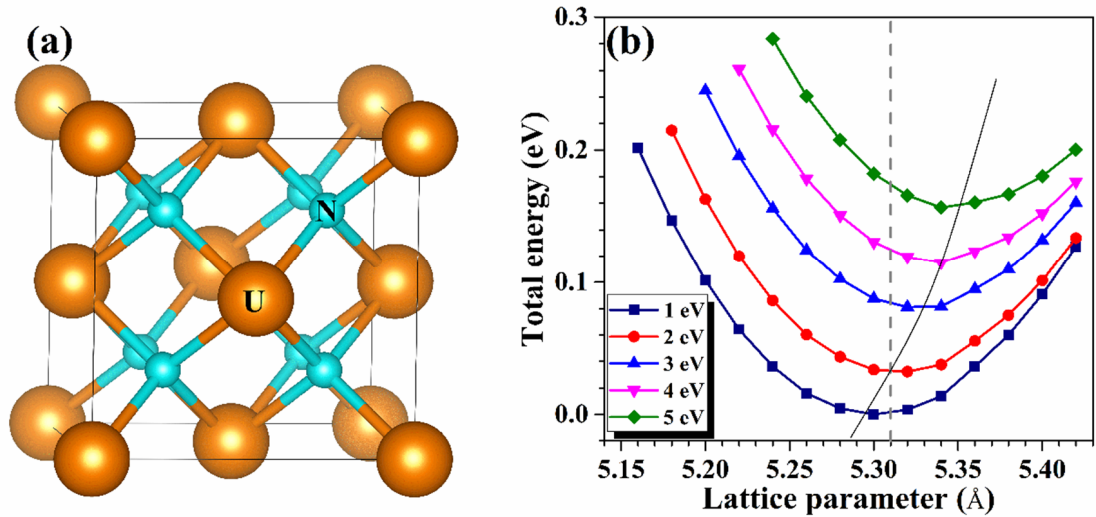

**Fig. 1.** (a) Ball-and-stick model of the $UN_2$ compound. (b) The total energy as a function of the lattice constant with Hubbard parameter $U$ ranged from 1 ~ 5 eV. The lowest energy for $U$ = 1 eV is set as 0 eV. The black trajectory shows the minimum energy and the vertical dashed line indicates experimental lattice constant.

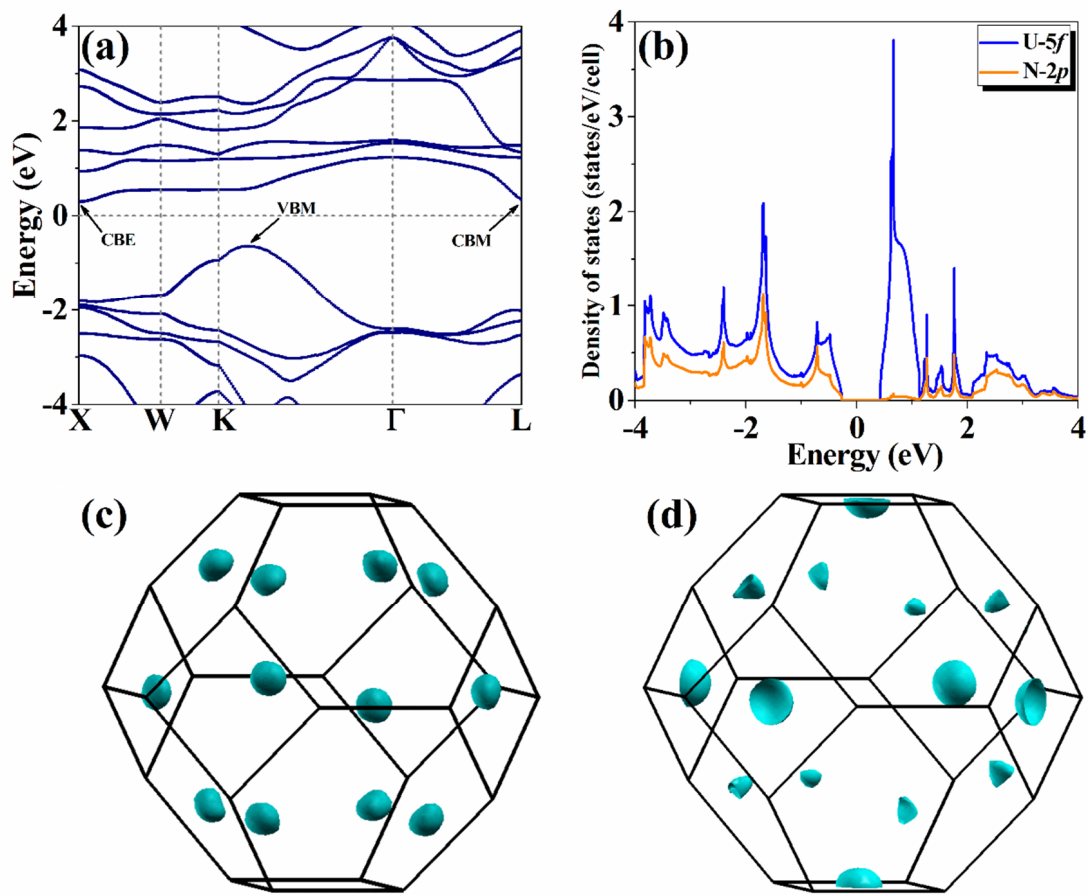

**Fig. 2.** (a) Band structures of the UN$_2$ calculated within DFT + $U$ + $GW$ method. (b) Orbital decomposed DOS of the UN$_2$. (c) and (d) are the isoenergy surfaces (0.1 eV from the band extremum) of the top valence band and bottom conduction band, respectively.

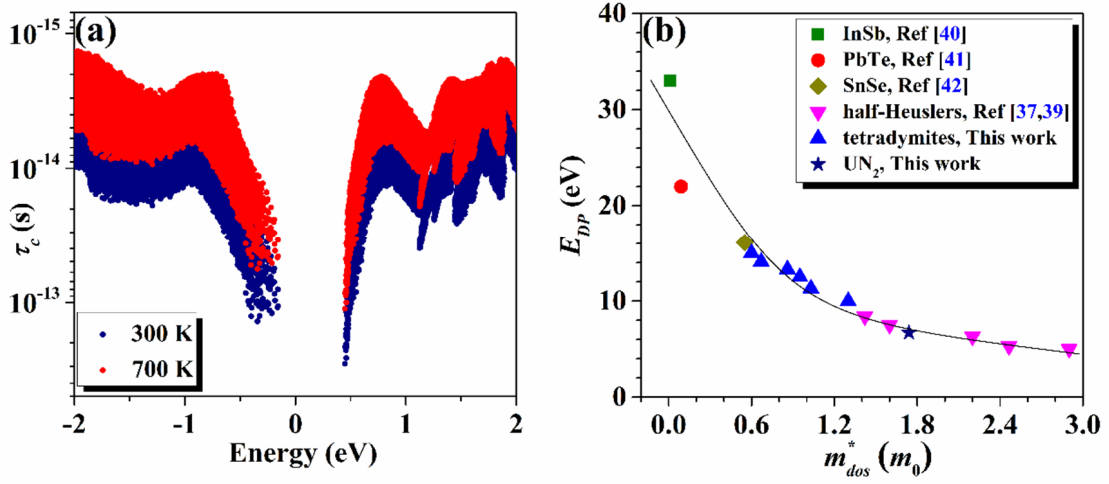

**Fig. 3.** (a) Energy-dependent carrier relaxation time of the $UN_2$ at 300 and 700 K. The Fermi level is at 0 eV. (b) The absolute value of deformation potential constant with respect to the DOS effective mass for the $UN_2$, as compared with some typical TE materials.

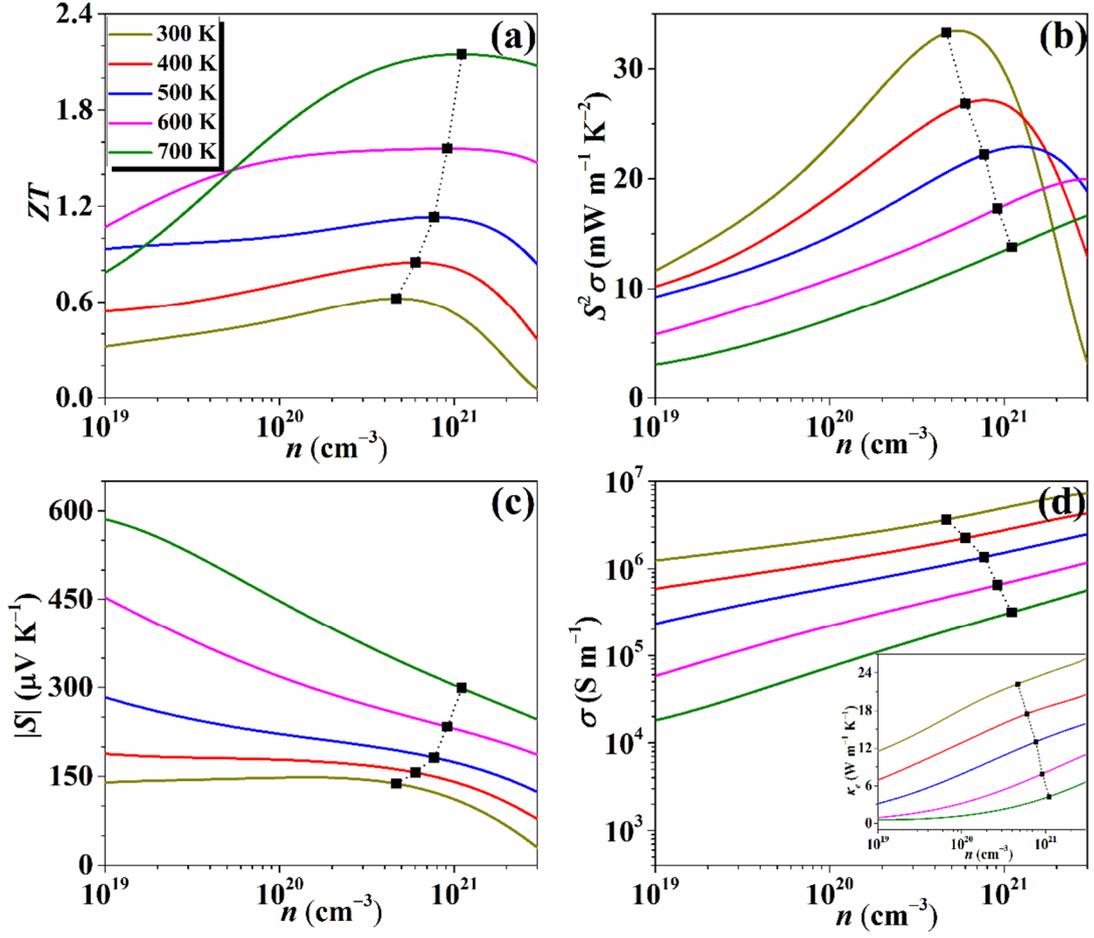

**Fig. 4.** (a) The *ZT* value, (b) the power factor, (c) the absolute Seebeck coefficient, and (d) the electrical conductivity of *n*-type UN$_2$, plotted as a function of carrier concentration at different temperature. The inset in (d) shows the corresponding electronic thermal conductivity.

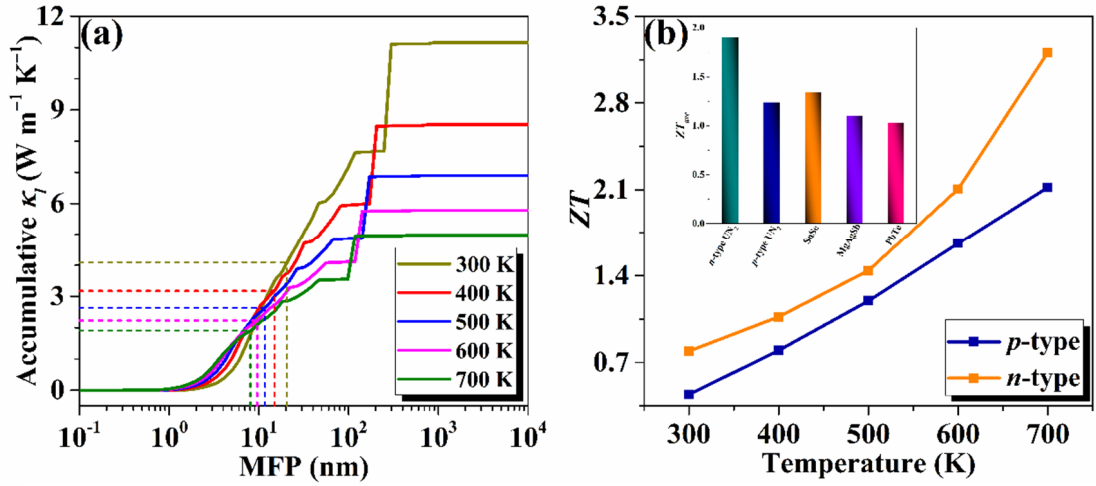

**Fig. 5.** (a) The accumulative lattice thermal conductivity of $UN_2$ with respect to the phonon MFP from 300 to 700 K. (b) The *ZT* value as a function of temperature when the lattice thermal conductivity is reduced by nanostructuring. The inset in (b) compares the $ZT_{ave}$ of the $UN_2$ with those of several representative TE materials.